\documentclass[12pt]{article}

\usepackage{graphicx}
\usepackage{braket}
\usepackage{color}
\usepackage{dsfont}
\usepackage{bm}
\usepackage{amsmath,amssymb}
\usepackage{hyperref}
\usepackage{times}
\usepackage[superscript]{cite}

\newcommand{\SI}[2]{$#1 \, \mathrm{#2}$}
\newcommand{\head}[1]{\textbf{}}
\newenvironment{sciabstract}{%
\begin{quote} \bf}
{\end{quote}}

\newcounter{lastnote}

\title{Heterodyne Sensing of Microwaves with a Quantum Sensor}%

\author{
Jonas Meinel, $^{1,2 \ast}$ Vadim Vorobyov, $^1$ Boris Yavkin,$^{1}$ Durga Dasari, $^{1,2}$ \\ Hitoshi Sumiya,$^3$ Shinobu Onoda,$^4$ Junichi Isoya,$^5$ J{\"o}rg Wrachtrup $^{1,2 \ast}$ \\
\\
\normalsize{$^{1}$3. Physikalisches Institut, IQST and Centre for Applied Quantum Technologies} \\
\normalsize{University of Stuttgart, Pfaffenwaldring 57, 70569 Stuttgart, Germany}\\
\normalsize{$^2$Max-Planck Institute for Solid State Research, Stuttgart 70569, Germany} \\
\normalsize{$^3$Advanced Materials Laboratory, Sumitomo Electric Industries Ltd., Itami 664-0016, Japan} \\
\normalsize{$^4$Takasaki Advanced Radiation Research Institute, National Institutes}\\
\normalsize{for Quantum and Radiological Science and Technology, Takasaki 370-1292, Japan} \\
\normalsize{$^5$ Faculty of Pure and Applied Sciences, University of Tsukuba, Tsukuba 305-8573, Japan.}\\
\normalsize{Correspondence: $^\ast$ j.meinel@pi3.uni-stuttgart.de, j.wrachtrup@pi3.uni-stuttgart.de}
}

\date{}
\begin{document}
\maketitle
\begin{sciabstract}
Diamond quantum sensors are sensitive to weak microwave magnetic fields resonant to the spin transitions.
However the spectral resolution in such protocols is limited ultimately by sensor lifetime. 
Here we demonstrate a heterodyne detection method for microwaves (MW) leading to a lifetime independent spectral resolution in the GHz range. 
We reference the MW-signal to a local oscillator by generating the initial superposition state from a coherent source. 
Experimentally we achieve a spectral resolution below \SI{1}{Hz} for a \SI{4}{GHz} signal far below the sensor lifetime limit of kilohertz.
Furthermore we show control over the interaction of the MW-field with the two level system by applying dressing fields, pulsed Mollow absorption and Floquet dynamics under strong longitudinal radio frequency drive. While pulsed Mollow absorption leads to highest sensitivity, the Floquet dynamics allows robust control independent from the systems resonance frequency.
 Our work is important for future studies in sensing weak microwave signals in wide frequency range with high spectral resolution.


\end{sciabstract}

\maketitle


\section{\label{sec:level1}Introduction}

\head{The problem of MW sensing} 
Precise detection of microwave frequency fields is of importance for a wide range of applications in cosmology \cite{Thornton2013}, radar \cite{Lloyd2008,Barzanjeh2020}, quantum optics with quantum circuit systems \cite{Bozyigit2011, Gasparinetti2017,Dmitriev2019} and electron spin signals or coupling to phonons \cite{MacCabe2019}. 
Oscillating magnetic fields, described by their amplitude and frequency, require a sensor with high sensitivity and high spectral resolution over a wide range of frequencies.

\head{How NV is better than other possible systems}
Atomic systems, such as Nitrogen vacancy (NV) centers in diamond, offer a microwave (MW) sensing platform through the electron spin transition \cite{joas2017quantum, Chipaux2015, Stark2017}.
Further the spin state can be optically pumped leading to an effective sensor temperature of about \SI{10}{mK} ($99\%$ polarization) \cite{joas2017quantum}, hence in principle being able to resolve single photon level signals \cite{Haikka2017}.
\head{Remaining problems with any Quantum systems}
However for any quantum sensors, the spectral resolution and sensitivity are linked through the lifetime of the system \cite{staudacher2013nuclear,aslam2017nanoscale,rosskopf2017quantum}. 
\head{State of the art, in other fields}
Advancements in experimental control of dynamical decoupling sequences allowed to separate sensitivity from spectral resolution \cite{schmitt2017submillihertz,boss2017quantum} widely applied for nuclear magnetic resonance (NMR) detection using NV-centers, e.g. \cite{Glenn2018, Bucher2020}. While with these techniques radio frequency signals can be detected with a spectral resolution below \SI{1}{Hz}, this technique fails for frequencies beyond \SI{10}{MHz}. 

\begin{figure}
\begin{center}
\includegraphics[scale = 0.5]{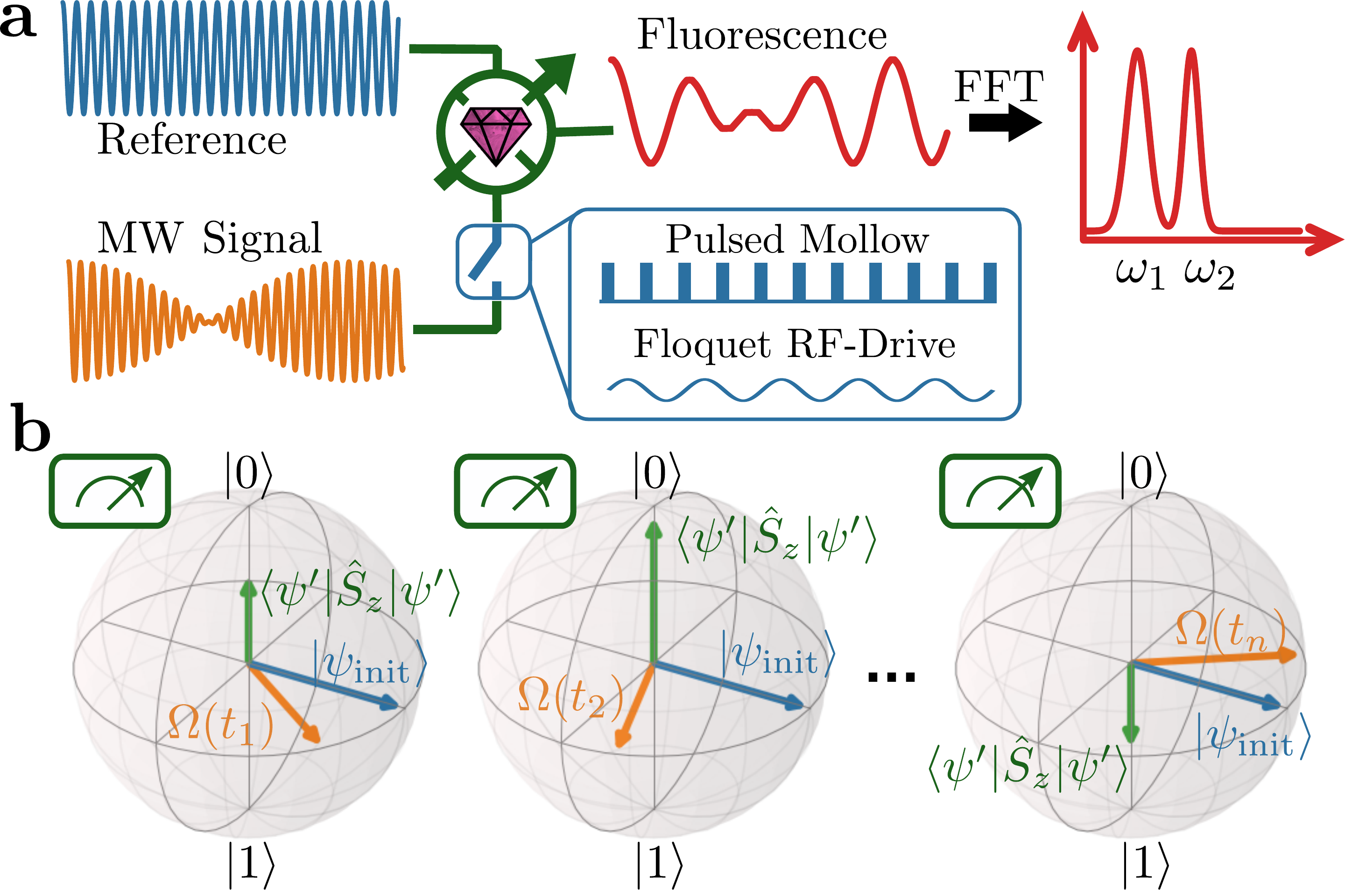}
\caption{\label{fig:motivation} \textbf{a} Measurement of a multi mode microwave frequency signal in a heterodyne scheme relative to an external frequency reference using the NV center as a sensor. The interaction with the signal is controlled with application of dressing fields. We benchmark two techniques: 1) pulsed Mollow absorption using dynamical decoupling sequence and 2) Floquet dynamics under strong RF-driving. The demodulated signal leads to high spectral resolution. \textbf{b} Heterodyne detection by creating the same initial state $|\psi_{\rm{init}}\rangle$ from a coherent external MW-source between sequential measurements. A long coherent signal $\Omega(t)$ is stroboscopically observed in the rotating frames. Finally the projection of $|\psi^\prime \rangle$, the final state after evolution under $\Omega$, is measured.}
\end{center}
\end{figure}

\head{Our contribution in solving the problem}
In our work we overcome this limitation and extend it in principle to the full MW spectrum \SI{0-100}{GHz}.
We created a heterodyne sensor illustrated in figure \ref{fig:motivation}, by mixing an external reference with the MW signal and detecting a demodulated signal in the fluorescence of the NV-center. 
\head{Overview of the paper}
This is achieved by creating the initial state of the sensor using above mentioned coherent external reference MW source. 
This state evolves under the signal field, and is sensitive to the relative phase between reference and signal. 
We further show that the interaction with the signal field can be controlled with dressing fields creating side bands. 
First we studied pulsed Mollow absorption, a dynamical decoupling sequence applied to sense MW-fields, making our protocol compliant with high sensitivity detection schemes and still achieving sensor unlimited spectral resolution. 
Second we study Floquet dressed states under strong longitudinal RF-drive allowing to create detection side bands independent from the resonance frequency of the quantum system making it applicable over a large frequency range.\newline

\section{Results}
\textbf{Theory}\\
In the following we theoretically describe how one achieves heterodyne sensing using a quantum sensor, here with NV-spins. The task at hand is to sense an oscillating microwave fields with frequency $\omega$, given by:
\begin{equation}
    \Omega(t,\phi_0) = \Omega_0 \cos(\omega t + \phi_0).
\end{equation}
$\Omega_0 = \gamma B_{\rm{signal}}$ is the amplitude of the field with $\gamma$ being the electron gyromagnetic ratio, and $ \phi_0$ is the initial phase of the signal. The coupling of such a field to the electron spin of the NV center, is simply given by the Hamiltonian $H = \vec{\Omega}(t)\vec{\hat{S}}$, where $\vec{\hat{S}}$ is the effective two-level system of the triplet ground-state spin configuration of the NV center. In secular approximation the overall Hamiltonian becomes:
\begin{equation}
    H = \omega_s \hat{S_z} +\Omega  \cos(\omega t + \phi_0) \hat{S_x},
\end{equation}
where $\omega_s$ is the transition frequency in the two level subspace of the NV-spin triplet, $\hat{S}_z = (\sigma_z - \mathds{1})/2$ and $\hat{S}_x = \sigma_x /\sqrt{2}$, where $\sigma$ are the pauli matrices and $\mathds{1}$ is the identity. After transformation into the rotating frame we get:
\begin{equation}
    H' = \frac{\Delta\omega}{2}\hat{\sigma_z} + \frac{\Omega}{2\sqrt{2}} \cos( \phi_0) \hat{\sigma_x} + \frac{\Omega}{2\sqrt{2}}\sin( \phi_0) \hat{\sigma_y},
\label{eq:hamiltonian_prime}
\end{equation}
where $\Delta \omega = \omega_s - \omega$, depicted e.g. as orange arrow in figure \ref{fig:motivation}b. 
To introduce a heterodyne approach, we have to add a second microwave field $\Omega_{\rm{ref}}(t,\phi_{\rm{ref}})$ acting as a reference. 
In our concept the reference source recreates the initial state with a $\pi/2$ pulse:
\begin{equation}
    \ket{\psi_{\rm{init}}(\phi_{\rm{ref}})} = (\ket{0} + \exp(i \phi_{\rm{ref}}) \ket{-1})/\sqrt{2},
\end{equation}
which evolves under the Hamiltonian in equation \ref{eq:hamiltonian_prime} for the time $\tau$:
\begin{equation}
    \ket{\psi(\tau)} = U(H'(\phi_0),\tau)  \ket{\psi_{\rm{init}}(\phi_{\rm{ref}})}.
\end{equation}
After evolution we measure the expectation value $\langle \hat{S_z}\rangle$:
\begin{equation}
    \langle S_z(\tau, \phi_{\rm{ref}}, \phi_0)\rangle \approx \Omega \tau \sin(\phi_0 - \phi_{\rm{ref}} )
\end{equation}
 for $\Omega'\tau, \frac{\Delta \omega}{\Omega'} \ll 1$ where $\Omega' = \sqrt{\Delta\omega^2 + \Omega^2}$. 
 With this measurement outcome, we now achieved a heterodyne response as we compare the phases of two microwave fields through the measurement of the $\hat{S}_z$ expectation value. 
 Further, we study a series of consecutive measurements recorded at time $t_{n+1} = t_{n} + T$, where $T$ is sampling interval. This leads to $\phi_{0,n+1} = \phi_{0,n} + \omega T$ and $\phi_{\rm{ref},n+1} = \phi_{\rm{ref},n} + \omega_{\rm{ref}}T$. Experimentally and theoretically it is instructive to analyze the auto correlation between the single measurement outcomes $S_n = S_z (t_n)$:
\begin{equation}
\label{eq:auto-corr}
\begin{split}
    C(n) &= \langle S_{n'} S_{n'+n} \rangle \\
    &    = \sum^M_{n'=1} \Omega^2 \tau^2 \sin(\delta\omega T n' + \delta\phi) \sin(\delta\omega T (n'+n)+ \delta\phi)\\
    &\approx \frac{1}{2} M \Omega^2 \tau^2 \cos(\delta \omega T n)
\end{split}
\end{equation}
where $\delta \omega = \omega_{\rm{ref}} - \omega$ is the demodulated frequency and $\delta \phi = \phi_{\rm{ref}} - \phi_0$ is the initial phase difference, which averages to zero under the auto correlation. While in this expression the auto correlation of the $S_z$ operator presented, in the experiment we acquire the auto correlation of the fluorescence photon counts collected in each measurement during the spin readout of NV center. These correlation functions are linked through the fluorescence spin contrast and and average fluorescence of the NV center \cite{pfender2019high}.

 In the discussion above we assumed a segmented evolution, starting with the state preparation by the reference followed by the free evolution under the signal field. While this approach shows clearly the heterodyne principle, in reality, a MW field is constantly interacting with the sensor. To achieve tunable interaction we further apply dressing fields.  This  disables the sensing field's influence during sensor state preparation ($\Omega/\Delta \omega \ll 1$) and switches it on during the interaction when the dressing field is applied ($(\Delta \omega - \Omega_{\rm{dressing}})/\Omega \ll 1$). 
 Here $\Omega_{\rm{dressing}}$ is the energy shift due to the dressing field, see methods \ref{app:mollow} and \ref{app:floquet} respectively for the analytical derivation. This leads to a Hamiltonian as described in equation \ref{eq:hamiltonian_prime}.\\
 \newline

\begin{figure*}

\includegraphics[width = \textwidth]{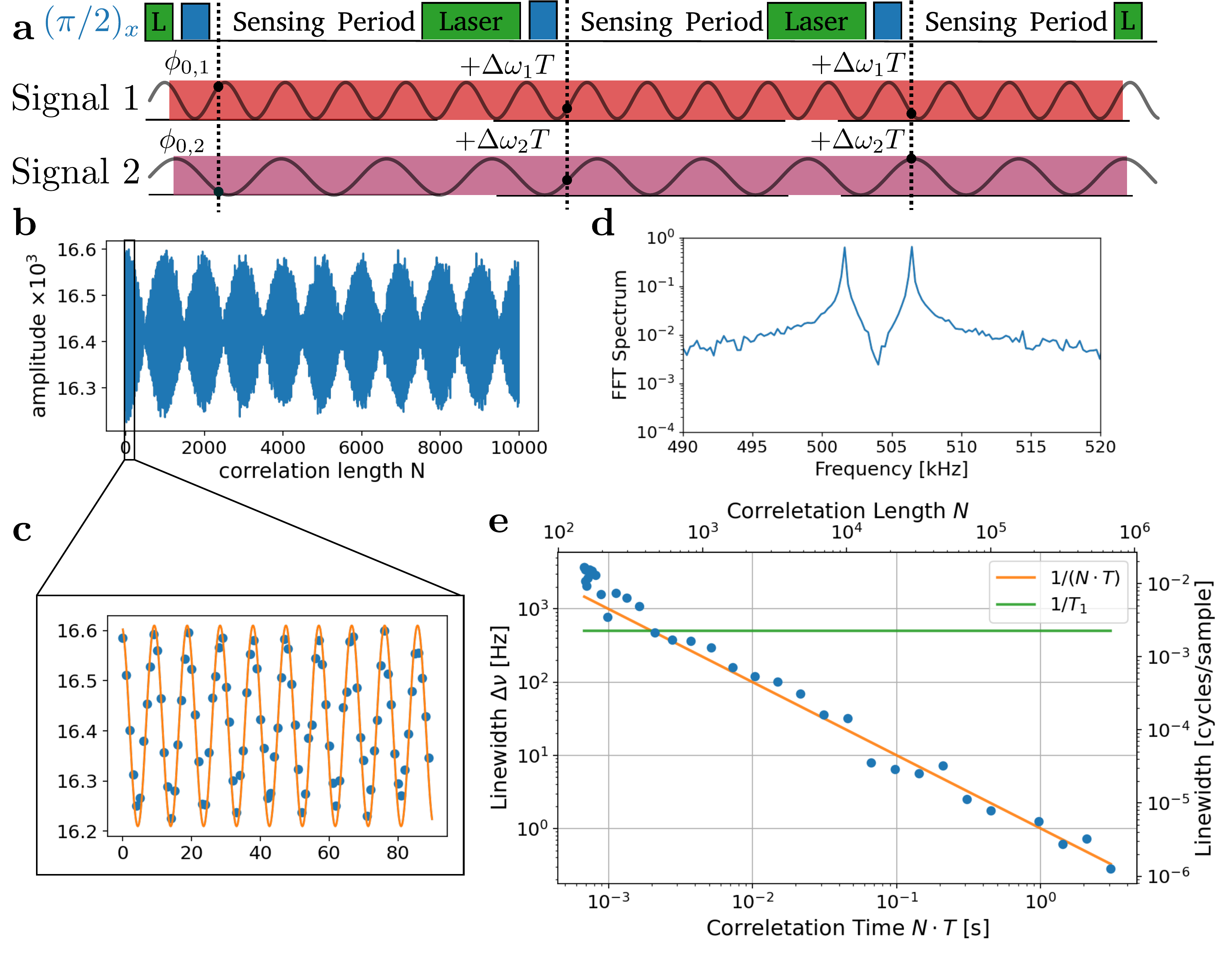}
\caption{\label{fig:spectral-resolution} \textbf{a} Heterodyne detection scheme for microwave fields by phase dependent evolution in the rotating frame. The phase of the external source evolves relative to the signal 1 ($\Delta\omega_1 = 2\pi\cdot  50.138 $ kHz) and signal 2 ($\Delta\omega_2 = \Delta\omega_1 + 2\pi \cdot 482 $ Hz) with $\Delta \omega_{1,2} \Delta t$ between two measurements. In \textbf{b} auto-correlation of the photon counter time trace. The beating between the two signals can be clearly recognized, consisting of fast oscillations shown in \textbf{c}. \textbf{d} Fourier spectrum of the auto-correlation. In \textbf{e} linewidth of the FFT peak as a function of the correlation length. The $1/(N \cdot T)$ scaling of the linewidth shows the Fourier limited linewidth for the applied sinusoidal signal. For the longest correlation time of 3 seconds we achieve a linewidth of 300 mHz, well below the sensor lifetime of $1/T_1 = 500$ Hz.}
\end{figure*}

\begin{figure}
\begin{center}
\includegraphics[scale = 0.5]{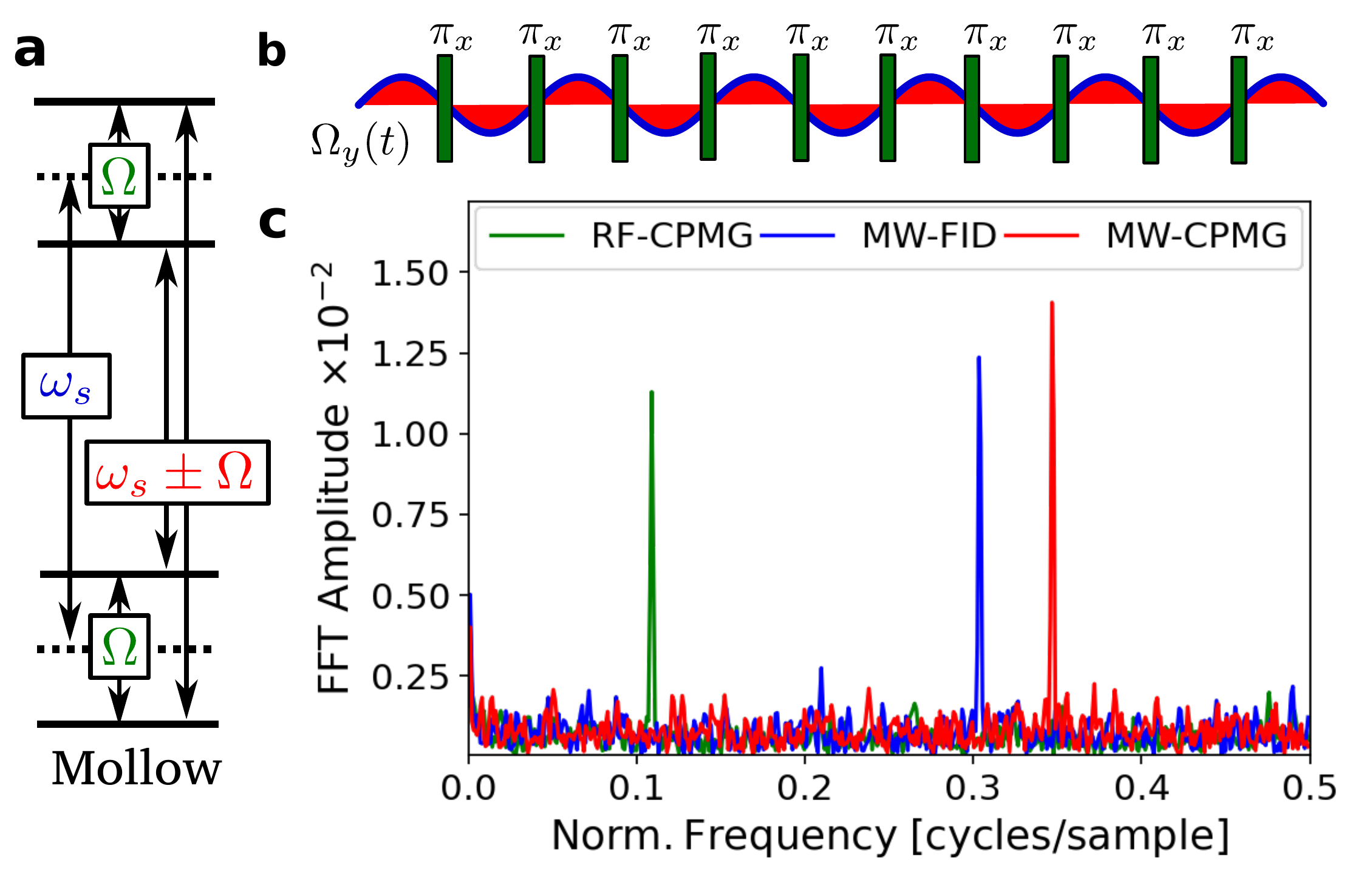}
\caption{\label{fig:summmary} Heterodyne detection is compatible with dynamical decoupling sequences. \textbf{a} A decoupling sequence creates from the initial two level system, with transition $\omega_s$, a Mollow triplet with two detection sidebands at $\omega_s \pm \Omega$, where $\Omega = \frac{\pi}{\tau}$, and a low frequency transition $\Omega$. A MW-field resonant to such a sideband is rotating in the spins reference frame. \textbf{b} Phase sensitive measurements of the rotating components are measured by applying a decoupling sequence. \textbf{c} Demodulated signal spectrum of the heterodyne measurement with the Mollow triplet transitions. Blue: resonant transition $ \omega_s$, green: RF-frequency $\Omega$ and red: Mollow sideband $\omega_s \pm \Omega$.}
\end{center}
\end{figure}

\begin{figure}
\begin{center}
\includegraphics[scale = 0.6]{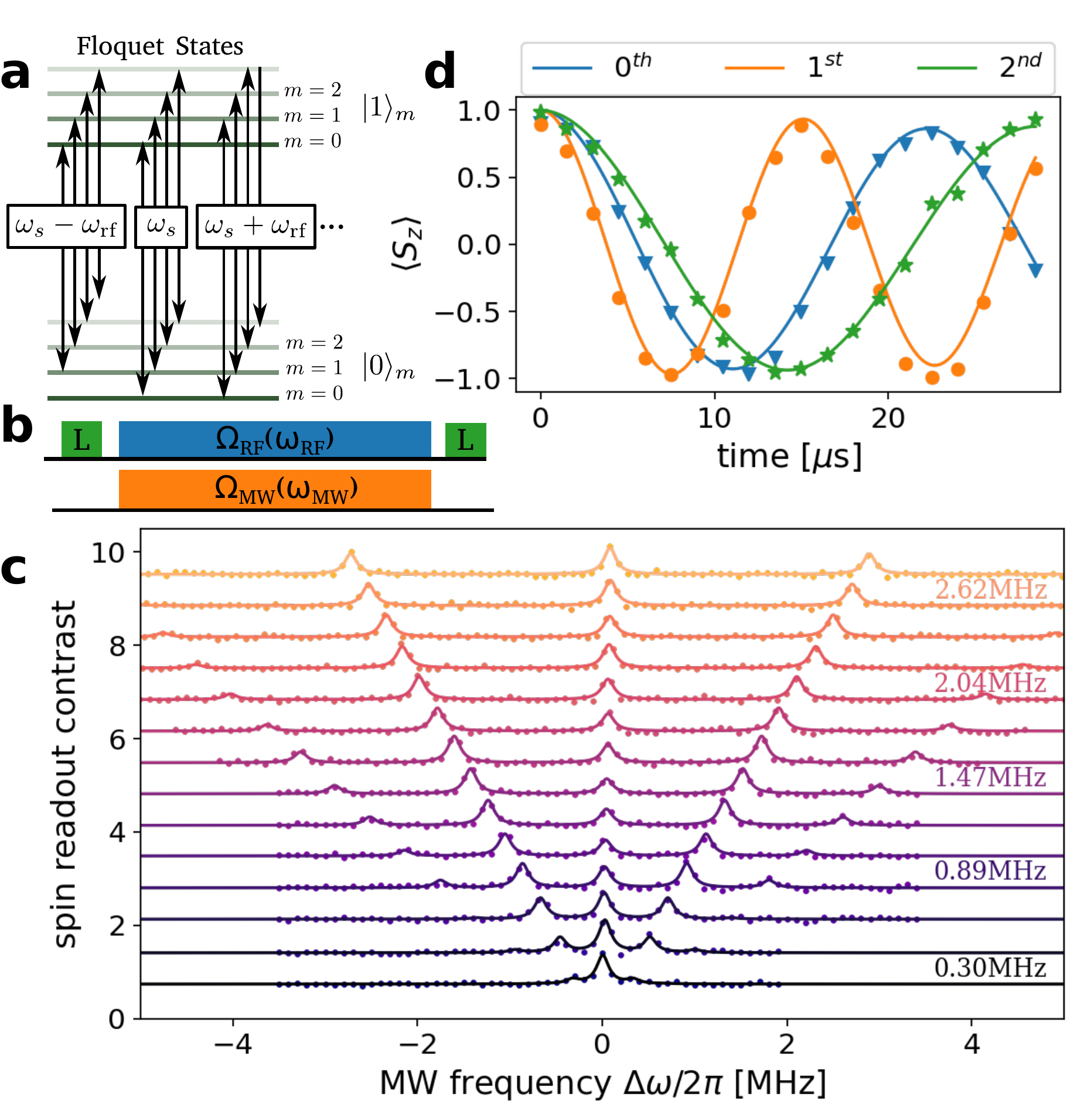}
\caption{\label{fig:floquet-rabi} The heterodyne detection of MW-fields with RF dressed system. \textbf{a} The RF driving creates the new energy states $E_{0,m} = m \omega_{\rm{rf}}$ and $E_{1,m} = \omega_s + m \omega_{\rm{rf}}$. Generating side bands with $\omega = \omega_s + \Delta m\ \omega_{\rm{rf}}$. \textbf{b} Experimental scheme for probing these transitions. \textbf{c} Optical detected magnetic resonance spectrum with increasing $\omega_{\rm{rf}}$ showing the Floquet sidebands at multiple of $\omega_{\rm{rf}}$. \textbf{d} Rabi oscillations at the central peak (0th), first and second sideband. The point chosen such that the Rabi frequency is largest for the first sideband and is described with $\nu_{\rm{Rabi,n}} =J_n(\frac{2.5 \rm{MHz}}{1.45 \rm{MHz}}) \cdot 0.125\ \rm{MHz}$.}
\end{center}
\end{figure}

\textbf{Heterodyne Detection of Microwave Fields}\\
We realize the heterodyne measurements by the scheme shown in figure \ref{fig:spectral-resolution}a. The NV spin is initialized in the $m_s = 0$ state with a green laser pulse, and the superposition state $\ket{-_i} = \frac{1}{\sqrt{2}} (\ket{0} - i \ket{-1})$ is created by applying a $\left(\frac{\pi}{2}\right)_x$ pulse from a coherent MW-source. The phase of the coherent source defines the initial state and the sensed signal interacts with the spin relative to it. Therefore the evolution is determined by the relative frequency difference of the signal with respect to the rotating frame $\delta \omega $ and the initial relative phase $\phi_{0}$ between the signal and the reference. 
In a proof of principle experiment we separate the state preparation from the interaction by applying the signal only during the sensing period for a time $\tau$. 
A series of measurements allows us to coherently measure the phase of the signal at the start of each measurement. Because the spectral resolution in heterodyne measurements is given by the reference and does not depend on the measurement time by $(1/\tau)$, it allows us to sense a multimode signal in parallel, illustrated as a second signal in the lower panel. 
In the experiment we measured a coherent two-frequency MW signal with frequencies $\omega_1 = 2\pi \cdot (4139.4 + 5 \cdot 10^{-2}) \, \rm{MHz}$, and $\omega_2 = 2 \pi \cdot (4139.4 + 5 \cdot 10^{-2} + 5 \cdot 10^{-4}) \, \rm{MHz}$. With respect to the reference (spin transition) frequency $\omega_s = 2\pi \cdot 4139.4 \, \rm{MHz}$, they are off-resonant by $\Delta \omega_1 =  50.138 \, \rm{kHz}$ and $\Delta \omega_2 =  50.620 \, \rm{kHz}$ respectively. Both signals have equal amplitudes of $\Omega_1 = \Omega_2 = 2 \pi \cdot 3.6 \;$MHz and are applied for $34.2 \,$ns. When we perform a series of $10^6$ measurements we record on average 0.1 photons per measurement. We extract the demodulation signal from the auto correlation, as defined in equation \ref{eq:auto-corr} and shown in figure \ref{fig:spectral-resolution}b. In the auto-correlation we clearly see the beating of the two signals and the enlarged view in figure \ref{fig:spectral-resolution}c shows the sinusoidal oscillations expected from the theoretical derivation above. We demonstrate the high spectral resolution by taking the fast Fourier transform  (FFT) of the auto correlation signal, as shown in figure \ref{fig:summmary}d. We observe a splitting of the two peaks corresponding to the frequency difference of ($\omega_1 - \omega_2 = 2 \pi\cdot 482 \, \rm{Hz}$). The narrow linewidth, of just \SI{1}{ Hz}, clearly demonstrates a separation beyond the coherence time of the sensor. Furthermore we analyze the Fourier peak in figure \ref{fig:spectral-resolution}e and find that the linewidth is Fourier limited by the correlation length ($1/(N T)$ scaling). For a maximal computed correlation length of 3 seconds, for a measurement of $\Delta \omega =2\pi\cdot 75$ kHz, $T = 1.824 \, \mu$s and $\Omega = 2 \pi \cdot 111 $ kHz, we achieve a \SI{300}{mHz} linewidth, a 3 orders of magnitude improvement compared to the $1/T_1\approx 0.5 \; \rm{kHz}$ lifetime limit, equivalently we are now sensitive to sub ppb (parts per billion) changes in the frequency of applied fields. While this demonstrates the high spectral resolution for MW-sensing, below we  investigate the control of the interaction of the spin with the MW-field.\\
\newline
\textbf{Dynamical decoupling increases sensitivity}\\
In the discussion above we applied the signal only during the sensing time. Here we introduce the pulsed Mollow triplet, created by dynamical decoupling, which effectively gives control over the interaction with the signal. 
Such decoupling sequences create a new dressed basis with new energy eigenstates shown in figure \ref{fig:summmary}a. While in most sensing applications of decoupling sequences the low frequency transition $\Omega$ is studied, we aim at sensing microwave fields and hence consider the Mollow sidebands at the absorption frequencies $\omega_s \pm \Omega$ \cite{joas2017quantum,stark2017narrow}. This allows to control the interaction of the sensor with the signal field. On top of that during the decoupling sequence the sensor lifetime increases up to $T_{1\rho}$, which is by orders of magnitude larger than $T_2^*$ \cite{Abobeih2018}. 
It is essential to have a precise reference frequency, which is achieved best when $\Omega$ is independent of the power of the dressing fields. Here we analyze a pulsed Mollow scheme which creates sidebands at $\Omega = \pi/\tau$, with $\tau$ the inter pulse spacing. This scheme is robust to power fluctuations (i.e. fluctuations in $\Omega$), which typically challenges continuous wave Mollow methods. 
The interaction of the pulse train with the MW-signal field is schematically drawn in figure \ref{fig:summmary}b, the $\pi$-pulses result in an integration of the rotating component of the MW-signal and therefore creating a maximal phase pickup of $2\Omega_{\rm{Signal}} T/\pi$, with $T$ the total sensing time. It becomes apparent that the phase of the oscillations relative to the pulse train alters the spin evolution. It is this effect which makes it suitable for heterodyne detection. Depending on the phase of the perpendicular component we get a transition from $\ket{+}$ to $\ket{-}$ across the poles. Experimentally we modified the scheme of figure \ref{fig:spectral-resolution}a where we now apply the pulse train during the sensing time. In figure \ref{fig:summmary}c we show the measured correlation result as Fourier transformation. We applied the Carr-Purcell-Meiboom-Gill-Sequence (CPMG) as a dynamical decoupling sequence, with 6.8 $\mu$s between two pulses and 10 repetitions leading to a total sensing time of 68 $\mu$s. We experimentally observe a Fourier limited linewidth for the Mollow peak analog to above with the exception that the applied signal was 15 times smaller. As a comparison we also show the low frequency transition $\Omega$, the well studied heterodyne peak for NMR, and the resonant absorption described in the section above. We see that all transitions in the dressed level scheme can be sensed in our heterodyne scheme and therefore relative to an external reference. With this treatment we created controllable transitions with increased sensitivity for heterodyne detection of microwaves. 
\newline
\\
\textbf{Tuning of Interactions with Floquet States}\\
Until now we only considered control over the sensor using MW-manipulation. 
In the following we consider an alternative approach by dressing the sensor states with a strong longitudinal RF-drive. The advantage is, that we naturally create side bands depending on the frequency of the RF-field and hence making it robust to amplitude fluctuations without the requirement of a pulsed operation and more importantly the power requirements for the MW-channel relax, which becomes particular relevant for high MW frequencies. While RF-control over the two level system has been studied  \cite{childress2010multifrequency,ashhab2007two}, we use it as a resource for heterodyne detection. The Hamiltonian under such RF-field driving with frequency $\omega_{\rm{rf}}$ and strength $\Omega_{\rm{rf}}$ becomes:
\begin{equation}
    H = \frac{\omega_0}{2} \sigma_z + \Omega_{\rm{rf}} \cos(\omega_{\rm{rf}} t) \sigma_z + \Omega_0 \sin(\omega t + \phi_0) \sigma_x,
 \end{equation}
where we consider a strong driving $\Omega_{\rm{rf}} \gtrapprox \omega_{\rm{rf}}$. The strong RF-drive results in new energy levels depicted in figure \ref{fig:floquet-rabi}a. The state $\ket{0}$ becomes $\ket{0,m}$ with a new quantum number, similar to the phonon excitations, this results into new energy eigenvalues of $E_{0,m} = m \omega_{\rm{rf}}$ and $E_{1,m} = \omega_s + m \omega_{\rm{rf}}$ respectively. 
These states allow transitions with frequency $\omega_s \pm \Delta m \, \omega_{\rm{rf}}$. Because of the energy splitting given by the frequency of the RF-drive, it becomes clear that these sidebands are ideal candidates for heterodyne detection. The transitions strength of the sidebands are given by $P_{\Delta m} =  J_{\Delta m}\left({\Omega_{\rm{rf}}}/{\omega_{\rm{rf}}}\right)$, and hence one could tune the transition strength by adjusting $x={\Omega_{\rm{rf}}}/{\omega_{\rm{rf}}}$ to be at a local maximum. At this local maximum we have a quadratic dependence on power fluctuations $(\Delta\Omega_{\rm{rf}})^2$ which underlines the robustness of the method. Experimentally, see fig. \ref{fig:floquet-rabi}b, we probe these new transition frequencies by initializing the sensor spin with a green laser pulse, followed by a simultaneously applied strong RF-field and a weak MW-probe field. The state is finally read out with another laser pulse. The optically detected magnetic resonance (ODMR) of the RF-dressed states is shown in figure \ref{fig:floquet-rabi} where we measure the spin readout contrast using single shot readout, see reference \cite{neumann2010single,Steiner2010}, with changing MW-probe frequency for RF-driving frequencies between $0.3-2.9 \, \rm{MHz}$ illustrated by a vertical offset. Finally we picked the driving frequency of \SI{1.45}{MHz} and performed Rabi oscillations on the central peak ($0^{th}$), first and second sideband, to confirm the dependence on the transition strengths $J_{\Delta m}(x) \Omega_{\rm{Rabi}}$, where $\Omega_{\rm{Rabi}}/2\pi =125 \, \rm{kHz}$. We measured $\Omega_0(x)/2\pi =45 \, \rm{kHz}$, $\Omega_1(x) = 66 \, \rm{kHz}$ and $\Omega_2(x) = 35 \, \rm{kHz}$ resulting into a value of $x = 1.72$, which is close to the maximum of $J_1(x)$.  

\begin{figure*}
\includegraphics[width = 0.98 \textwidth ]{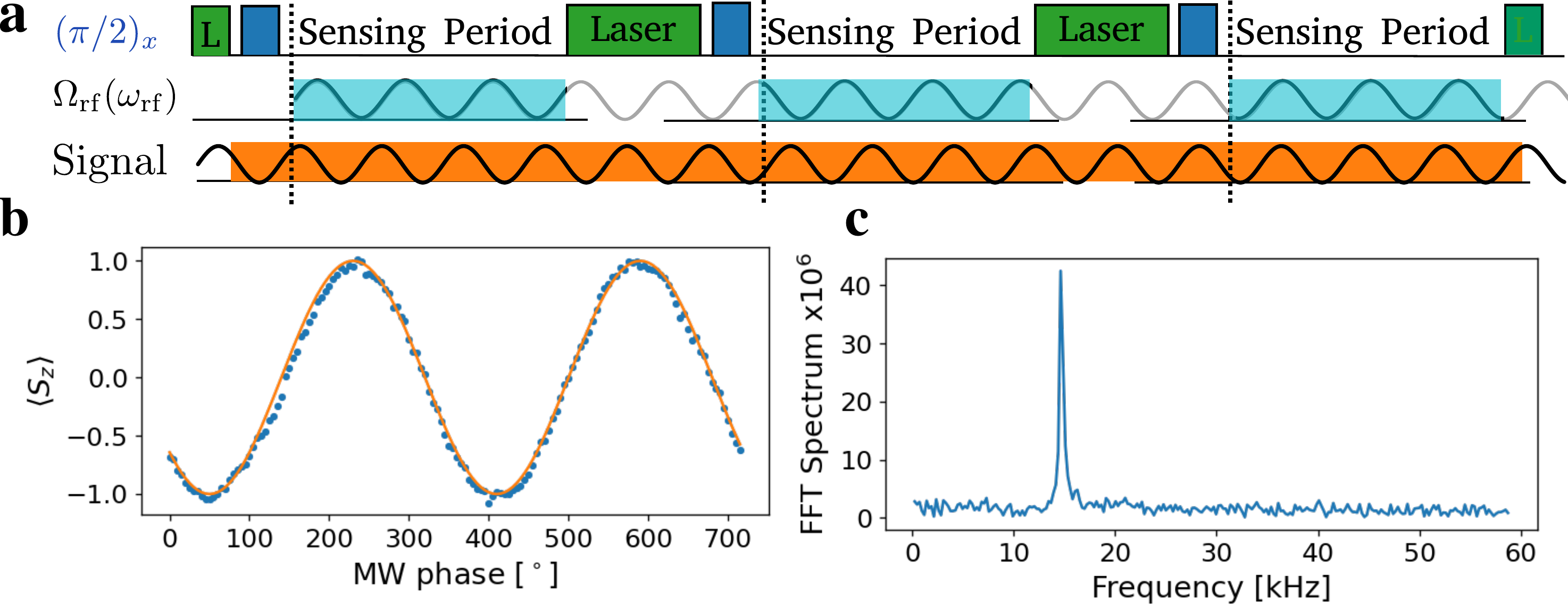}
\caption{ \label{fig:floquet-qudyne} Heterodyne detection using Floquet dressed states. \textbf{a} Experimental scheme: Initialization and readout of the electron spin. Additional a RF-field is applied for the sensing period, which is coherent between measurements. The demodulation frequency is given by $\Delta\omega - \omega_{\rm{rf}}$ for the first sideband. \textbf{b} Sensor response as a function of the signal phase in the Floquet dressed system, while the RF phase is fixed. \textbf{c} Heterodyne detection of the MW-field in the Floquet dressed system. The phase of the RF-field is modulated by $\phi_{i+1} = \phi_{i} + 45^\circ$, moving the zero demodulation frequency to $\nu_{\rm{demod.}}$ = 1/8 measurements for resonant measurements.}
\end{figure*}

After introducing the new dressed states we show that they are suitable for heterodyne detection (for a detailed derivation of the phase sensitivity see the appendix). 
The RF-driving results in additional oscillations of the spin in the rotating frame around the z-axis and therefore can be interpreted as z-phase gates. Coming back to the picture presented for pulsed mollow absorption (figure \ref{fig:summmary}b) we have, instead of x-rotations, z-rotations which follow the frequency of the RF-drive. 
Depending on the phase of the RF-drive the timing of these gates changes and create a phase sensitive sequence. Higher harmonics appear because of saturation of the phase gates, surpassing $2\pi$ rotation within one period, creating higher order frequency acceptance.
Experimentally we show heterodyne detection in the RF-dressed basis by the scheme shown in figure \ref{fig:floquet-qudyne}a. We prepare the initial state of NV electron spin with a laser pulse and a $\pi/2$ pulse from a coherent source yielding a $\ket{-_i}$ state. Additionally we apply strong RF-drive during the sensing period. The signal interferes with the spin only during the sensing period giving us control over the interaction and we expect a demodulated frequency of $\omega_{\rm{signal}} - \Delta m \,\omega_{\rm{rf}} - \omega_s$. We experimentally show the phase sensitivity in figure \ref{fig:floquet-qudyne}b by repeating one sensing block, with varying initial phase of the MW-signal. We observe clean sinusoidal oscillations illustrating the phase sensitivity clearly. Finally we performed sequential measurements of a MW-field and plot the Fourier spectrum of the auto-correlation. The experimental conditions were adjusted such that demodulated frequency was zero. To see the oscillation we varied the initial phase of the RF-field depending on the measurement index as $\phi_{n+1} = \phi_n + 45^\circ$ moving the zero frequency to 1/8 of the sampling frequency. The peak width is Fourier limited. \newline
In summary we implemented a heterodyne sensing protocol for MW-frequencies, first with a simple phase sensitive free evolution in the rotating frame leading to lifetime unlimited resolution, below \SI{1}{Hz}. We further improved the sensitivity of this heterodyne scheme with pulsed Mollow absorption from $T_2^*$ to $T_{1,\rho}$, leading to estimated sensitivities of $5 \,\rm{nT}/\sqrt{\rm{Hz}}$ for single NV centers and $1\, \rm{pT}/\sqrt{\rm{Hz}}$ for an ensemble of NV centers\cite{joas2017quantum}. Finally we applied the RF-driven states to heterodyne sensing allowing us to control the interaction with the signal and creating stable sidebands, independent of the frequency band of the two level system. 

\section{Discussion}

This work extends heterodyne sensing with a quantum sensor to high frequencies up to \SI{100}{GHz}, not accessible with previous protocols \cite{boss2017quantum, schmitt2017submillihertz}. At the same time it converts absorption based MW sensing to a heterodyne scheme, extending thus the resolution beyond $T_1$ of the sensor \cite{joas2017quantum, Stark2017}. 
In this work we further show how to convert an absorption measurement protocol to a heterodyne detection protocol in three systems: a two level system, a Mollow triplet, and a Floquet dressed system. 
The key requirements for such conversion is to make a system sensitive to the phase of the incident signal and benchmark this on the single electron spin of NV center in diamond.

While the simplest case of two level system shows the working principle, the addition of Floquet dressing makes the interaction with the sensor controllable, and the Mollow dressing improves the sensitivity from $T_2^*$ to $T_{1,\rho}$ limited.

In this work we overcome the spectral resolution problem existing for quantum sensors of microwave signals, which is of importance for sensing weak signals with highest possible spectrum resolution for highly coherent microwave signals, e.g. maser radiation, quantum radar \cite{Lloyd2008,Barzanjeh2020} and Doppler velocimetry technologies, weak cosmic radiation or wireless communication protocols. 
Additionally, the heterodyne approach in sensing leads to the concept of sequential weak measurements of quantum systems, which potentially is important in measuring the quantum behaviour of mesoscopic bosonic or fermionic systems at high frequencies and for quantum feedback \cite{Martin2019}. 
Due to general nature of our work it could be applied to other two level systems, such as transmon qubits, which are naturally suitable for the task of microwave radiation sensing \cite{Bozyigit2011, Gasparinetti2017,Dmitriev2019} similar to what was shown for Rydberg atoms \cite{Jing2020}.


\newpage

\section{Acknowledgments}
We are grateful to Minsik Kwon for helpful
comments on the manuscript. JM acknowledges the Max Planck Graduate Center for Quantum Materials, MPI IMPRS and the IRTG research program. We acknowledge financial support by the German Science Foundation (the DFG) via SPP1601, FOR2724, the European union via ASTERIQS, and the European Research Council (ERC) under the European Union’s Horizon 2020 research and innovation programme (grant agreement No. 742610 SMel), the Max Planck Society, the Volkswagen Stiftung, IRTG GRK 2198/1, MPG IMPRS.

\section{Author contributions} 
J.M., V.V., B.Y. and J.W. conceived and designed the experiment. J.M., V.V. and D.D. developed the theory. H.S., S.O. and J.I. prepared the sample. J.M., V.V. and J.W. wrote the manuscript with inputs from all others. All authors read and commented on the manuscript.\\
\textbf{Competing Interests:}
The authors declare no competing interests.
\textbf{}
\bibliographystyle{naturemag}
\bibliography{apssamp}

\appendix

\section{Methods}

\subsection{Experimental Setup}
\begin{figure}
\begin{center}
\includegraphics[scale = 0.6]{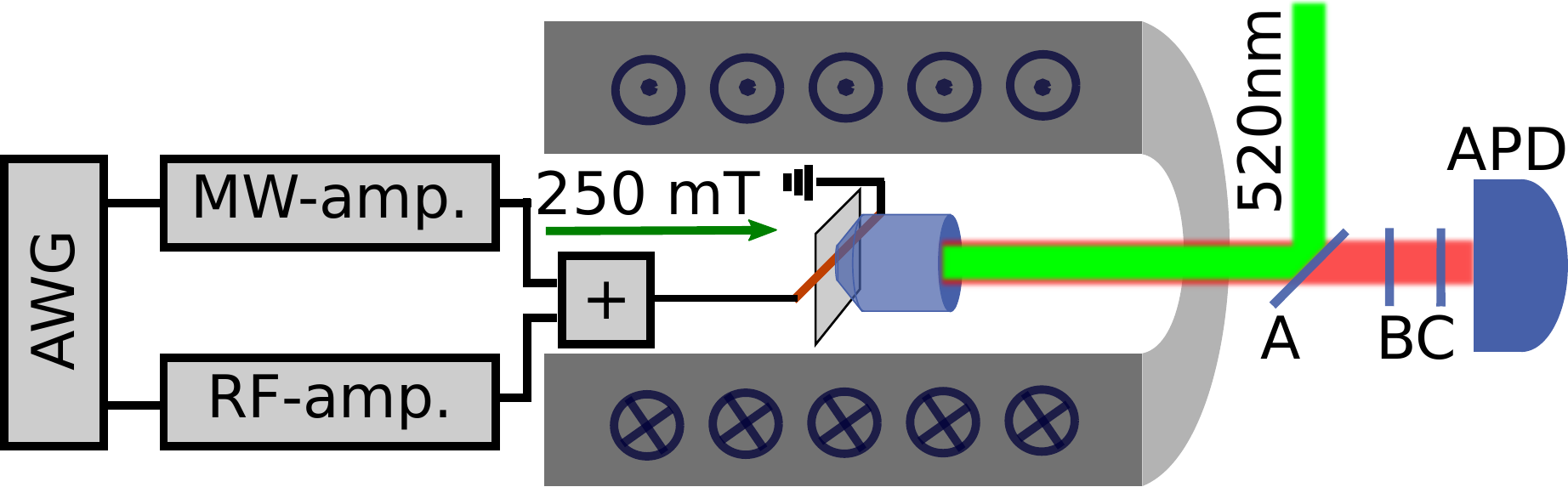}
\caption{\label{fig:setup} The diamond with individual NV-centers is studied with a confocal microscope in the center of a room temperature bore of a superconducting magnet, operated at \SI{250}{mT}. For excitation a \SI{520}{nm} diode laser is used, the beam is reflected from a dichroic mirror (A) and coupled into the objective. The fluorescence of the NV is passing again through the dichroic mirror (A), through a pinhole \SI{50}{\mu m}, (B) and finally through a long pass filter, \SI{650}{nm}, (C) before getting detected with an avalanche photon detector (APD) (Perkin-Elmer SPCM) . The microwave and radio frequencies are generated on a two channel arbitrary waveform generator, sent to amplifiers and combined before the microwave structure,  where the diamond is glued on.}
\end{center}
\end{figure}

The scheme of the experimental setup is shown in figure \ref{fig:setup}. The diamond crystal is positioned within a superconducting NMR-magnet (Scientific Magnetics). The magnetic field was about \SI{250}{mT} and aligned with NV center quantization (Z) axis, leading to transition frequency of \SI{4139.3}{MHz} between $\ket{0}$ and $\ket{-1}$ for the nuclear spin projection $m_I = +1$. For the optical excitation and collection of the fluorescence of the NV center we use an immersion oil objective with a numeric aperture of 1.35 and detect it with an avalanche photo-diode(APD) (Perkin-Elmer SPCM), capable of detecting single photons. A \SI{520}{nm} diode laser is used for excitation, which can be directly turned on/off within \SI{10}{ns}. The NV microwave transitions and RF-manipulation is generated on a Keysight 8190A arbitrary waveform generator with \SI{12}{GSamples/s}. The same device controls the laser diode switching and the data acquisition triggering. We amplify the first channel (MW) up to about \SI{40}{dBm} power with Hughes-TWT 8010H amplifier (max. \SI{7}{MHz} Rabi frequency), the second channel (RF) to about \SI{52}{dbm} with RF amplifier (Amplifier Research 150A250) leading to an oscillating field of \SI{0.1}{mT} along the NV quantization axis. We combine the MW and RF channels before coupling to a coplanar waveguide.
The crystal is a \SI{99.995}{\%} $^{12}$C-enriched diamond crystal (\SI{5.3\times 4.7 \times 2.6}{mm^3}) was grown by the temperature gradient method under high-pressure and high-temperature. A polished, (111)-oriented slice (\SI{2 \times 2 \times 0.08}{mm^3}) has been used in the present work.  The coherence times of the NV center studied here are $T_2^*=50 \, \mu s$ and $T_2\approx 300 \, \mu s$. The further characteristics of the sample and setup can also be found in reference \cite{pfender2019high}.

\subsection{Evolution under phase dependant Hamiltonian}
The essence of our work is the phase dependant measurement outcome. In the following we derive this outcome from a the protocol shown in figure \ref{fig:spectral-resolution}. We consider the following phase dependant Hamiltonian : 
\begin{equation}
    H' = \frac{\Delta\omega}{2}\sigma_z + \frac{\Omega}{2} \cos( \phi_0) \sigma_x + \frac{\Omega}{2}\sin( \phi_0) \sigma_y,
\end{equation}
When we create the time evolution operator $U$ we get:
\begin{equation}
\begin{split}
    U =& \exp(-iH' t) \\
      =& \exp\left(-\frac{i}{2}\Omega' t (\vec{n} \cdot\vec{\Sigma}) \right)\\
     =& \cos(\Omega' t) \mathds{1} - i \frac{\sin(\Omega' t)}{{\Omega^\prime}} \\
     & \times \left( \Omega \cos(\phi_0) \sigma_x + \Omega \sin(\phi_0) \sigma_y + \Delta \omega \sigma_z \right) \\
\end{split}
\end{equation}{}
where the $\vec{\Sigma} = (\sigma_x, \sigma_y, \sigma_z)$ is the Pauli vector and $\Omega' = \sqrt{ \Omega^2 + \Delta\omega^2}$ is the generalized Rabi frequency.
We assume an initial state of $\ket{+} = \frac{1}{\sqrt{2}}(\ket{0} + \ket{1})$, leading to the following state after evolution: 
\begin{equation}
\begin{split}
    \ket{\psi(t)}  = &\left( \cos(\Omega't) - i\frac{\sin(\Omega't)}{\Omega'} \Omega_x\right) \ket{+}\\ 
    +& \left(\frac{\sin(\Omega't)}{\Omega'}\Omega_y - i \frac{\sin(\Omega't)}{\Omega'} \Delta \omega\right)  \ket{-}
\end{split}
\end{equation}
where $\Omega_x = \Omega \cos \phi_0, \, \Omega_y =  \Omega \sin \phi_0$.
\begin{equation}
\begin{split}
    \langle S_z\rangle &= \sin(2\Omega' t)\frac{\Omega}{2\Omega'}\sin(\phi_0) + \sin^2(\Omega't)\frac{\Omega \Delta\omega}{\Omega'^2}\cos(\phi_0)
    \\
    &\approx \Omega t \sin(\phi_0),
\end{split}
\end{equation}
for small phase acquisition.\\

\subsection{Analytical Solution for Mollow Dressed System}
\label{app:mollow}
For control of the interaction with the signal field and long coherence times we apply dressed states, the Mollow triplet. The Hamiltonian in the lab frame is given by:
\begin{equation}
    H = \frac{\omega_0}{2} \sigma_z + G \sigma_x \cos(\omega_0 t) + \gamma \sigma_x \cos(\omega_1 t + \phi)
\end{equation}
When we go into the rotating frame of the fast oscillations $\omega_0$ we get:
\begin{equation}
    H' = (\frac{G}{2} + \frac{\gamma}{2} \cos(\Delta \omega t + \phi)) \sigma_x + \frac{\gamma}{2}\sin(\Delta \omega t + \phi) \sigma_y,
\end{equation}
ignoring the term $2 \,\omega \gg G$. Finally we can go into the second rotating frame:
\begin{equation}
    H'' = \Delta'' \sigma_x + \frac{\gamma}{4} (\cos\phi \, \sigma_z - \sin \phi \, \sigma_y) 
\end{equation}
where $\Delta'' = \frac{1}{2} (G - \Delta \omega + \gamma \cos(\Delta \omega t + \phi)) \approx \frac{1}{2}(G - \Delta \omega)$ and we assume $2 \Delta \omega \gg \Delta''$.\\

\subsection{Analytical Solution in the Strongly Driven System with Longitudional RF field}
\label{app:floquet}
This derivation was adapted from \cite{ashhab2007two} to our situation. We start with the Hamiltonian of the driving:
\begin{equation*}
    H = \frac{\omega_s}{2}\sigma_z +\frac{\Omega_{\rm{rf}}}{2} \cos(\omega_{\rm{rf}} t) \sigma_z + \Omega_1 \cos(\omega_{\rm{mw}}t+\phi_0)\sigma_x
\end{equation*}
Which we have to transform in the rotating frame with an evolution operator:
\begin{equation}
    U = \exp\left(\frac{i\sigma_z}{2} \left(\omega_st + \frac{\Omega_{\rm{rf}}}{\omega_{\rm{rf}}} \sin(\omega_{\rm{rf}} t)\right)  \right)
\end{equation}
The Hamiltonian in the rotating frame becomes:
\begin{equation}
    H^\prime =  \Omega_1 \cos(\omega_{\rm{mw}}t+\phi_0) \ U^\dagger\sigma_x U
\end{equation}{}
with $ \ U^\dagger\sigma_x U $ given by:
\begin{equation}
   \begin{pmatrix}
0 & \rm{e}^{-i\left(\omega_st + \frac{\Omega_{\rm{rf}}}{\omega_{\rm{rf}}} \sin(\omega_{\rm{rf}} t)  \right)}\\
\rm{e}^{i\left(\omega_st + \frac{\Omega_{\rm{rf}}}{\omega_{\rm{rf}}} \sin(\omega_{\rm{rf}} t)  \right)} & 0 \end{pmatrix}
\end{equation}
We can further use:
\begin{equation}
    \exp\left(i z \sin(\omega t)\right) = \sum_{k=-\infty}^{\infty} J_k\left(z\right) \exp\left(k \omega t\right)
\end{equation}{}
which leads to the following matrix entries of $U^\dagger\sigma_x U$:
\begin{equation}
    \begin{pmatrix}
0 &  \sum J_k(\frac{\Omega_{\rm{rf}}}{\omega_{\rm{rf}}}) e^{-i(\omega_s +k \omega_{\rm{rf}}) t}\\
 \sum J_k(\frac{\Omega_{\rm{rf}}}{\omega_{\rm{rf}}}) e^{i(\omega_s +k \omega_{\rm{rf}}) t} & 0 \end{pmatrix}
\end{equation}
Let's assume that the probe field amplitude is small $\Omega_1 \ll \omega_{\rm{rf}}$, which allows us to only look at one resonance and the probe field is close to resonance of a sideband $\omega_k =   \omega_s + k \,\omega_{\rm{rf}}$. We get the effective Hamiltonian:
\begin{equation}
\begin{split}
    H_k^\prime = &J_k\left(\frac{\Omega_{\rm{rf}}}{\omega_{\rm{rf}}}\right)  \Omega_1 \cos(\omega_{\rm{mw}}t+\phi_0) \\
    & \times\left(\cos(\omega_k t) \sigma_x + \sin(\omega_k t) \sigma_y \right)
\end{split}
\end{equation}{}
Using the rotating wave approximation, because of  $\Omega_1 \ll \omega_{\rm{rf}}$, we further get:
\begin{equation}
\begin{split}
    H_k^\prime =\frac{\Omega_k}{2}
    \left(\cos(\Delta^\prime t + \phi_0) \sigma_x + \sin(\Delta^\prime t + \phi_0) \sigma_y \right)
\end{split}
\end{equation}
with $\Omega_k =  J_k\left(\frac{\Omega_{\rm{rf}}}{\omega_{\rm{rf}}}\right)  \Omega_1$ and $\Delta^\prime =  \omega_{\rm{mw}} -  \omega_n$, which finally becomes:
\begin{equation}
\begin{split}
    H_n^{\prime \prime}= \Delta^\prime   \sigma_z + \frac{\Omega_n}{2}
    \left(\cos \phi_0 \, \sigma_x + \sin \phi_0 \, \sigma_y \right)
\end{split}
\end{equation}
showing the dependants on the initial phase $\phi_0$.

\end{document}